# Frustrated antiferromagnetic spin chains of edge-sharing tetrahedra in volcanic minerals $K_3Cu_3(Fe_{0.82}Al_{0.18})O_2(SO_4)_4$ and $K_4Cu_4O_2(SO_4)_4Me$Cl

L. M. Volkova[1] and D. V. Marinin[1]

**Abstract** The calculation of the sign and strength of magnetic interactions in two noncentrosymmetric minerals (klyuchevskite, $K_3Cu_3(Fe_{0.82}Al_{0.18})O_2(SO_4)_4$ and piipite, $K_4Cu_4O_2(SO_4)_4Cu_{0.5}$Cl) has been performed based on the structural data. As seen from the calculation results, both minerals comprise quasi-one-dimensional frustrated antiferromagnets. They contain frustrated spin chains from edge-sharing $Cu_4$ tetrahedra with strong antiferromagnetic couplings within chains and very weak ones between chains. Strong frustration of magnetic interactions is combined with the presence of the electric polarization in tetrahedra chains in piipite. The uniqueness of magnetic structures of these minerals caused by peculiarities of their crystal structures has been discussed.

**Keywords:** Frustrated quasi-one-dimensional antiferromagnets, tetrahedral spin chains, klyuchevskite $K_3Cu_3(Fe_{0.82}Al_{0.18})O_2(SO_4)_4$, piypite $K_4Cu_4O_2(SO_4)_4Me$Cl.

## 1 Introduction

The technological developments put forward new requirements to properties of the applied crystalline materials. In this regard, the potential of crystals of simple chemical compounds are almost exhausted. Here, one of the ways of further development will consist in selection of materials with required properties from natural objects and creation of novel functional materials on their basis. The objective of the present work was to reveal magnetic materials, which can be of not only scientific, but also of practical importance, among fumaroles of Tolbachik volcanoes (Kamchatka Peninsula, Russia) [1].

The search and study of compounds with the spin $S = 1/2$, whose magnetic subsystem is strongly frustrated, prevents, in some cases, the formation of a long-range order until realization of exotic states such as 'spin ice' and 'spin liquid', have been constituted one of the key focuses in the physics of the condensed state during the recent 20 years [2-10]. The most active studies were concerned with frustrated magnetics composed of vertex-sharing tetrahedra of magnetic ions. It is well-known that this type of the most frustrated magnetic lattice can be found in spinels and compounds of the pyrochlore type. We suggest that minerals of fumaroles of Tolbachik volcanoes can be also strongly frustrated antiferromagnets. The point is that the framework of crystal structures of many of these minerals is formed by anion-centered $XM_4$ tetrahedra of magnetic ions, for instance, $[OCu_4]^{6+}$, linking to each other with formation of isle-like complexes, infinite chains, layers, or frameworks [11-14]. Therefore, these minerals crystal structure itself provides the possibility of the emergence of a strongly frustrated magnetic system. The authors from the St. Petersburg school of structural mineralogy and crystal chemistry determined the crystal structures, studied the crystal chemistry, and performed systematization of minerals with anion-centered complexes [11-14]. However, unlike structural properties, the magnetic properties of these minerals have been studied very poorly yet.

The objective of the present work was to find genetic basics determined by the crystal structure and determining, in its turn, the magnetic structure and properties of two noncentrosymmetric minerals: klyuchevskite ($K_3Cu_3(Fe_{0.82}Al_{0.18})O_2(SO_4)_4$ [15]) and piipite ($K_4Cu_4O_2(SO_4)_4Cu_{0.5}Cl$ [16]). For this purpose, based on the data on crystal structures of these minerals, characteristics (sign and strength) of magnetic interactions were calculated, and their competition was examined. These compounds are characterized with similar geometries of exchange bonds - quasi-one-dimensional chains from edge-sharing copper tetrahedra.

---

[1] Institute of Chemistry, Far Eastern Branch of Russian Academy of Sciences, 159, 100-Let Prosp., Vladivostok 690022, Russia
e-mail: volkova@ich.dvo.ru

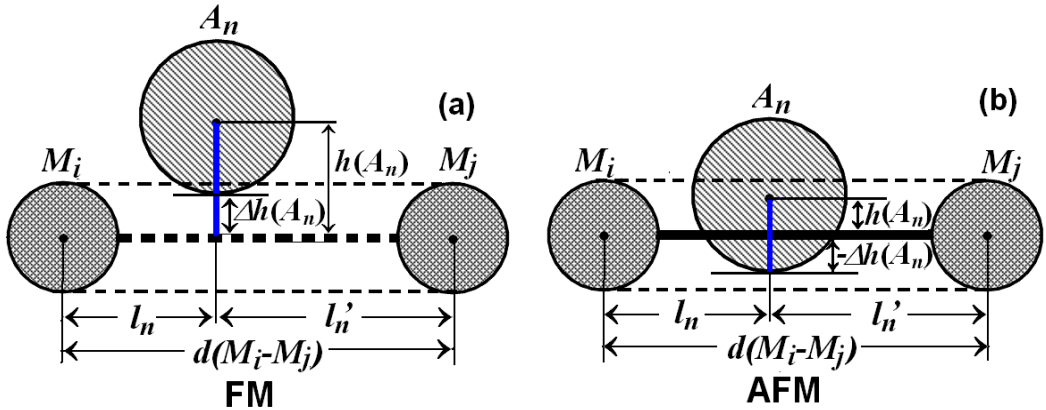

Fig. 1. A schematic representation of the intermediate $A_n$ ion arrangement in the local space between magnetic ions $M_i$ and $M_j$ in cases when the $A_n$ ion initiates the emerging of the ferromagnetic (a) and antiferromagnetic (b) interactions. $h(A_n)$, $l_n$, $l'_n$ and $d(M_i–M_j)$ - parameters determining the sign and strength of magnetic interactions $Jn$.

**2 Method of Calculation**

To determine the characteristics of magnetic interactions (type of the magnetic moments ordering and strength of magnetic coupling) in minerals $K_3Cu_3(Fe_{0.82}Al_{0.18})O2(SO_4)_4$ and $K_4Cu_4O_2(SO_4)_4MeCl$, we used the earlier developed phenomenological method (named the *"crystal chemistry method"*) and the program "MagInter" created on its basis.[17-21] In this method three well-known concepts about the nature of magnetic interactions are used. Firstly, Kramers's idea [22], according to which in exchange couplings between magnetic ions. separated by one or several diamagnetic groups, the electrons of nonmagnetic ions play a considerable role. Secondly, Goodenough–Kanamori–Anderson's model [23–26], in which crystal chemical aspect points clearly to the dependence of strength interaction and the type of orientation of spins of magnetic ions on the arrangement intermediate anions. Thirdly, as in polar Shubin–Vonsovsky's model [27], by consideration of magnetic interactions we took into account not only anions, which are valent bound with the magnetic ions, but also all the intermediate negatively or positively ionized atoms, with the exception of cations of metals with no unpaired electrons.

The method enables one to determine the sign (type) and strength of magnetic couplings on the basis of structural data. According to this method, a coupling between magnetic ions $M_i$ and $M_j$ emerges in the moment of crossing the boundary between them by an intermediate ion $A_n$ with the overlapping value of ~0.1 Å. The area of the limited space (local space) between the $M_i$ and $M_j$ ions along the bond line is defined as a cylinder, whose radius is equal to these ions radii. The strength of magnetic couplings and the type of magnetic moments ordering in insulators are determined mainly by the geometrical position and the size of intermediate $A_n$ ions in the local space between two magnetic ions $M_i$ and $M_j$. The positions of intermediate ions $A_n$ in the local space are determined by the distance $h(A_n)$ from the center of the ion $A_n$ up to the bond line $M_i$-$M_j$ and the degree of the ion displacement to one of the magnetic ions expressed as a ratio $(l_n'/l_n)$ of the lengths $l_n$ and $l_n'$ ($l_n \leq l_n'$; $l_n' = d(M_i-M_j)-l_n$) produced by the bond line $M_i$-$M_j$ division by a perpendicular made from the ion center (Fig. 1).

**Table 1** An estimate of $J_n$ magnetic couplings in oxides $Cu^{2+}$, $Fe^{3+}$ and $Cu^{2+}$-$Fe^{3+}$ by a crystal chemical method (I) and experimental and quantum-chemical methods (II)

| Compounds | Space group, lattice parameters | d(M-M) (Å) | $\|J\|$ (Å$^{-1}$) I (This work) | $J$, (degree Kelvin) II | $K^{(a)}$ | $K^a \times J$(Å$^{-1}$) (degree Kelvin) |
|---|---|---|---|---|---|---|
| AgCuVO$_4$, 293K [29] ICSD - 419201 | *Pnma* (No. 62): $a$ = 9.255, $b$ = 6.778, $c$ = 5.401Å, Z = 4 | d(Cu-Cu) = 3.389 | 0.0474 (AFM) | 330 (AFM) [30] | 6591 | 312 (AFM) |
| AgCuVO$_4$, 120K [29] ICSD – 419202 | *Pnma* (No. 62): $a$ = 9.242, $b$ = 6.775, $c$ = 5.396 Å, Z = 4 | d(Cu-Cu) = 3.388 | 0.0483 (AFM) | 330 (AFM) [30] | 6591 | 318 (AFM) |
| BaCu$_2$Ge$_2$O$_7$ [31] ICSD - 51282 | *Pnma* (No. 62: $a$ = 7.048, $b$ = 13.407, $c$ = 7.028 Å Z = 4 | d(Cu-Cu) = 3.546 | 0.0864 (AFM) | 540 (AFM) [32, 33] | 6591 | 569 (AFM) |
| Cu$_3$(AsO$_4$)(OH)$_3$ [34] ICSD - 68456 | $P2_1/c$ (No. 14): $a$ = 7.257, $b$ = 6.457, $c$ = 12.378 Å $\beta$ = 99.51º, Z = 4. | d(Cu-Cu) = 3.131 | 0.0532 (AFM) | 300 (AFM) [35] 423 (AFM) [35] | 6591 | 351 (AFM) |
|  |  | d(Cu-Cu) = 3.663 | 0.1337 (AFM) | 700 (AFM) [35] 884 (AFM) [35] | 6591 | 881(AFM) |
| Cu$_2$Te$_2$O$_5$Cl [36] ICSD- 89978 | $P\bar{4}$ (No. 81): $a$ = 7.621, $c$ = 6.320 Å, Z = 2 | d(Cu-Cu) = 3.230 | 0.0117 (AFM) | 38.5 (AFM) [36] 40.9 (AFM) [37] | 6591/2 | 38.6 (AFM) |
| Cu$_2$Fe$_2$Ge$_4$O$_{13}$ [38] ICSD - 240615 | $P2_1/m$ (No. 11): $a$ = 12.088, $b$ = 8.502, $c$ = 4.870 Å $\beta$ = 96.17º, Z = 2 | d(Cu-Cu) = 3.021 | 0.0812 (AFM) | 278.4 (AFM) [39-41] | 6591/2 | 268 (AFM) |
|  |  | d(Cu-Fe) = 3.072 | 0.0544 (AFM) | 27.8 (AFM) [39-41] | 511 | 27.8 (AFM) |
|  |  | d(Fe-Fe) = 3.208 | 0.0443 (AFM) | 18.6 (AFM) [39-41] | 399 | 17.3 (AFM) |
| KFe$_3$(SO$_4$)$_2$(OH)$_6$ [42], ICSD - 12107 | $R\bar{3}m$ (No. 166): $a$ = 7.315, $c$ = 17.224 Å, $\gamma$ = 120º, Z = 3 | d(Fe-Fe) = 3.658 | 0.0947 (AFM) | 37 (AFM) [43] 45 (AFM) [44, 45]] | 399 | 37.8 (AFM) |
| FeTe$_2$O$_5$Cl [46] ICSD-240492 | $P2_1/c$ (No. 14): $a$ = 13.153, $b$=6.595, $c$ = 14.145 Å, $\beta$ = 108.77º, Z = 8 | d(Fe-Fe) = 3.151 | 0.0330 (AFM) | 10.2 (AFM) [46] | 399 | 13.2 (AFM) |
|  |  | d(Fe-Fe) = 3.328 | 0.0249 (AFM) | 10.9 (AFM) [46] | 399 | 9.9 (AFM) |
| FeTe$_2$O$_5$Br [46] ICSD - 240490 | $P2_1/c$ (No. 14): $a$ = 13.396, $b$=6.597, $c$ = 14.290 Å, $\beta$ = 108.12º, Z = 8 | d(Fe-Fe) = 3.159 | 0.0361 (AFM) | 11.7 (AFM) [46] | 399 | 14.4 (AFM) |
|  |  | d(Fe-Fe) = 3.343 | 0.0420 (AFM) | 19 (AFM) [47] | 399 | 16.8 (AFM) |
| K$_4$Cu$_4$O$_2$(SO$_4$)$_4$Cu$_{0.5}$Cl [16] ICSD - 64684 | $I4$ (No. 79): $a$ = 13.60 Å, $c$ = 4.98 Å, Z=2 | d(Cu-Cu) = 2.936 | 0.0719 (AFM) |  | 6591/2 | 237 (AFM) |
|  |  | d(Cu-Cu) = 3.242 | 0.0689 (AFM) |  | 6591/2 | 227 (AFM) |
| K$_3$Cu$_3$(Fe$_{.82}$Al$_{.18}$)O$_2$(SO$_4$)$_4$ [15], ICSD - 67698 | $I2$ (No. 5): $a$ = 18.667, $b$ = 4.94, $c$ = 18.405 Å, $\beta$=101.5º, Z=4 | *tetrahedron I* |  |  |  |  |
|  |  | d(Cu-Cu) = 3.220 | 0.0986 (AFM) |  | 6591 | 650 (AFM) |
|  |  | d(Cu-Cu) = 2.899 | 0.0756 (AFM) |  | 6591 | 498 (AFM) |
|  |  | d(Cu-Cu) = 2.870 | 0.0604 (AFM) |  | 6591 | 398 (AFM) |
|  |  |  | (0.0028) (FM) |  | 6591 | 18.5 (FM) |
|  |  | d(Fe-Cu) = 2.918 | 0.0761[b] (AFM) |  | 511 | 38.9[b] (AFM) |
|  |  | d(Fe-Cu) = 3.400 | 0.0671[b] (AFM) |  | 511 | 34.3[b] (AFM) |
|  |  | d(Fe-Cu) = 3.364 | 0.0515[b] (AFM) |  | 511 | 26.3[b] (AFM) |
|  |  | *tetrahedron II* |  |  |  |  |
|  |  | d(Cu-Cu) = 3.331 | 0.0719 (AFM) |  | 6591 | 473.9[b] (AFM) |
|  |  | d(Cu-Cu) = 3.115 | 0.0463 (AFM) |  | 6591 | 305.2[b] (AFM) |
|  |  | d(Fe-Cu) = 3.268 | 0.0854[b] (AFM) |  | 511 | 43.6[b] (AFM) |
|  |  | d(Fe-Cu) = 3.193 | 0.0801[b] (AFM) |  | 511 | 40.93[b] (AFM) |
|  |  | d(Fe-Fe) = 4.940 | 0.0025[b] (AFM) |  | 399 | 1.0[b] (AFM) |

[a]Scaling factor for translating the value per angstrom into Calvin's degree in $Cu^{2+}$ and $Fe^{3+}$ oxides
[b]As accepted for $J_n$ calculations, the Fe position is fully occupied

The intermediate $A_n$ ions will tend to orient magnetic moments of $M_i$ and $M_j$ ions and make their contributions $j_n$ into the emergence of antiferromagnetic (AFM) or ferromagnetic (FM) components of the magnetic interaction in dependence on the degree of overlapping of the local space between magnetic ions ($\Delta h(A_n)$), asymmetry ($l_n'/l_n$) of position relatively to the middle of the $M_i$-$M_j$ bond line, and the distance between magnetic ions ($M_i$-$M_j$).

Among the above parameters, only the degree of space overlapping between the magnetic ions $M_i$ and $M_j$ ($\Delta h(A_n) = h(A_n) - r_{A_n}$) equal to the difference between the distance $h(A_n)$ from the center of $A_n$ ion up to the bond line $M_i$-$M_j$ and the radius ($r_{A_n}$) of the $A_n$ ion determined the sign of magnetic interaction. If $\Delta h(A_n) < 0$, the $A_n$ ion overlaps (by $|\Delta h|$) the bond line $M_i$-$M_j$ and initiates the emerging contribution into the AFM-component of magnetic interaction. If $\Delta h(A_n) > 0$, there remains a gap (the gap width $\Delta h$) between the bond line and the $A_n$ ion, and this ion initiates a contribution to the FM-component of magnetic interaction.

The sign and strength of the magnetic coupling $J_{ij}$ are determined by the sum of the above contributions:

$$J_{ij} = \sum_n j_n$$

The value $J_{ij}$ is expressed in $\text{Å}^{-1}$ units. If $J_{ij} < 0$, the type of $M_i$ and $M_j$ ions magnetic ordering is AFM and, in opposite, if $J_{ij} > 0$, the ordering type is FM.

The method is sensitive to insignificant changes in the local space of magnetic ions and enables one to find intermediate ions localized in critical positions, deviations from which would result in the change of the magnetic coupling strength or spin reorientation (AFM-FM transition, for instance, under effect of temperature or external magnetic field).

The format of the initial data for the "MagInter" program (crystallographic parameters, atom coordinates) is in compliance with the cif-file in the Inorganic Crystal Structure Database (ICSD) (FIZ Karlsruhe, Germany). The room-temperature structural data and ionic radii of Shannon [28] were used for calculations.

The comparison of our data with that of other methods shows that the scaling factors $K$ for translating the value in per angstrom into Calvin's degree in oxides $Cu^{2+}$ (spin-1/2, $Cu^{2+}$-$Cu^{2+}$), $Fe^{3+}$ (spin-5/2, $Fe^{3+}$-$Fe^{3+}$) and ($Cu^{2+}$-$Fe^{3+}$) are equal 6591x$n$ (where $n$ = Z/4, Z – cell formula units), 399 and 511, respectively (Table 1). Energy Converter: 1 degree Kelvin = 0.0862 meV.

Studies of the minerals of interest were performed in the following order:
- the sign and strength of all the magnetic interactions between magnetic ions as inside low-dimensional fragments of the sublattice of magnetic ions as between them were calculated;
- the probability of the emergence of anomalies of magnetic interactions and magnetic phase transitions in case of insignificant changes in the local space between magnetic ions was determined;
- the specific geometric configurations in sublattices of magnetic ions, in which the competition of magnetic interactions takes place, were identified.
- the conclusions on these compounds magnetic structures were made based on the obtained data on characteristics of magnetic interactions and the presence of geometric frustrations in these interactions.

Tables 2 and 3 (section 3) show the crystallographic characteristics and parameters of magnetic couplings ($J$n) calculated on the basis of structural data and respective distances between magnetic ions in the materials under study. Besides, for intermediate X ions providing the maximal contributions ($j^{max}$) into AFM or FM components of these $J$n couplings, the degree of overlapping of the local space between magnetic ions $\Delta h$(X), the asymmetry $l_n$'/$l_n$ of the position relatively to the middle of the $M_i$-$M_j$ bond line, and the $M_i$-X-$M_j$ angle are presented.

## 3 Results and Discussion

### 3.1 Klyuchevskite, $K_3Cu_3(Fe_{0.82}Al_{0.18})O_2(SO_4)_4$

Klyuchevskite ($K_3Cu_3(Fe_{0.82}Al_{0.18})O_2(SO_4)_4$)[15] crystallizes in the noncentrosymmetric monoclinic $I2$ system. Magnetic $Cu^{2+}$ ions occupy 3 crystallographically independent sites Cu1, Cu2, and Cu3 and have a characteristic distortion of $Cu^{2+}$ coordination polyhedra due to the Jahn-Teller effect strengthened by geometric hindrances related to the packing features. Elongated tetragonal pyramids $CuO_5$ with a strong shift of copper ions to basal planes serve as the coordination surrounding of Cu1, Cu2, and Cu3. As a result, the Cu-O distances to apical pyramid vertices (Cu1–O11 = 2.54 Å, Cu2–O12 = 2.47 Å, and Cu3–O8 = 2.86 Å) are significantly longer than those to basal vertices (1.70-2.05 Å). The Cu-O distances to the sixth oxygen atom exceed 2.95 Å for all copper ions. The $Fe^{3+}$ ions occupy the only crystallographically independent site (Fe1) and have the octahedral surrounding (Fe1–O = 1.78-2.07 Å) with an insignificant distortion, as compared to the $Cu^{2+}$ coordination. According to Ref. 15, in this klyuchevskite sample, 18 % of $Fe^{3+}$ ions are substituted by nonmagnetic $Al^{3+}$ ions. However, we consider its crystal structure as a model, in which the Fe1 position is fully occupied by magnetic $Fe^{3+}$ ions and assume that such a compound can be synthesized.

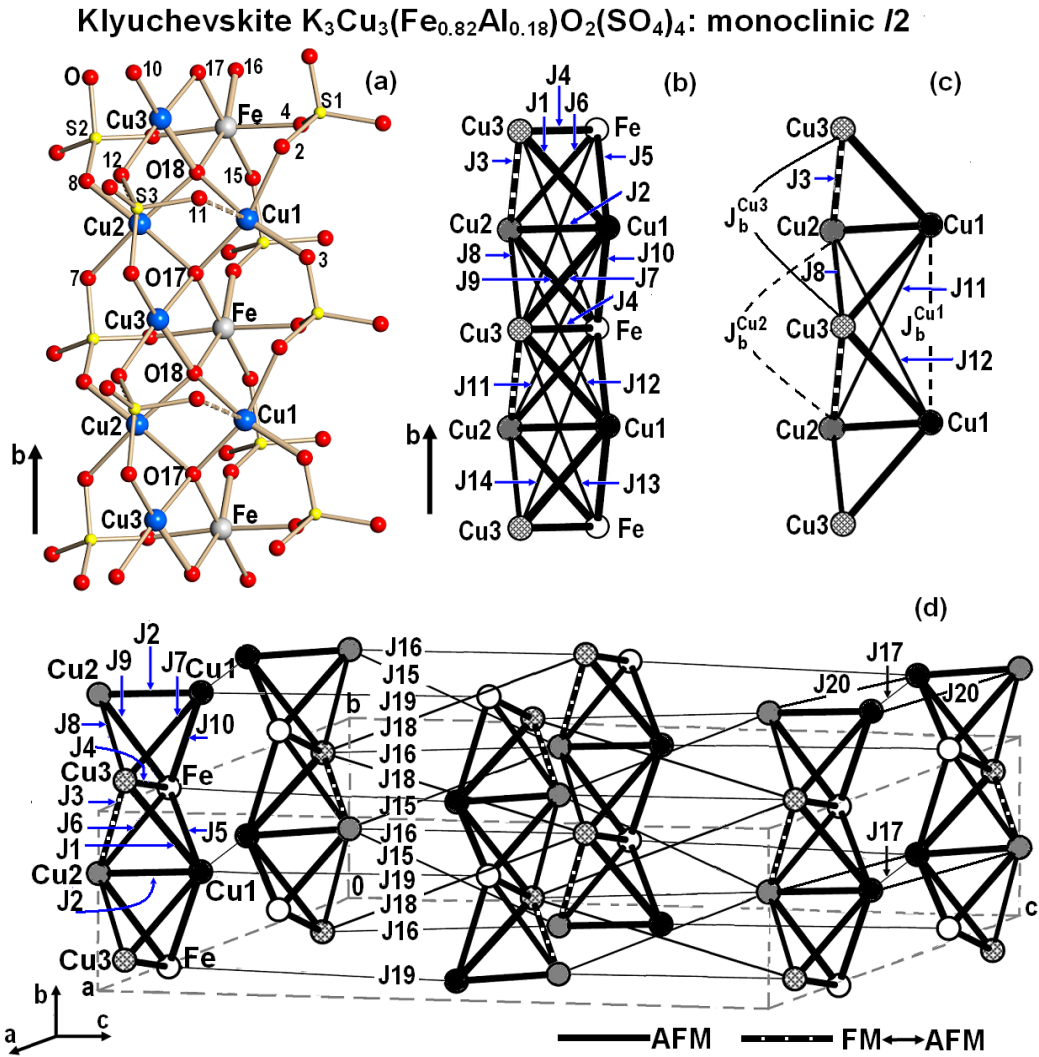

**Fig. 2** The chain $[Cu_3Fe^{3+}O_2(SO_4)_4]^{3-}$ in the klyuchevskite $K_3Cu_3(Fe_{0.82}Al_{0.18})O_2(SO_4)_4$ along the *b* axis (a). The sublattice of magnetic ions $Cu^{2+}$ and $Fe^{3+}$ and the coupling $J_n$ in klyuchevskite: (b) the chain of edge-sharing ($Cu_3Fe$) tetrahedra, (c) the corrugated chain of copper triangles alternately linked through side and vertex in case of $Fe^{3+}$ ions substitution by nonmagnetic ions, (d) intrachain and interchain $J_n$ couplings. In this and other figures the thickness of lines shows the strength of $J_n$ coupling. AFM and FM couplings are indicated by solid and dashed lines, respectively. The possible FM →AFM transitions are shown by the stroke in dashed lines.

The characteristic feature of this structure, as of a majority of crystal structures of minerals from volcanic exhalations of Kamchatka (Russia)[11-15], consists in the presence of complexes of anion-centered $OMe_4$ tetrahedra (Fig. 2a, b). 'Extra' oxygen anions (O17 and O18) 'pull over' magnetic

**Table 2**. Crystallographic characteristics and parameters of magnetic couplings (*J*n) calculated on the basis of structural data and respective distances between magnetic ions in klyuchevskite $K_3Cu_3(Fe_{0.82}Al_{0.18})O_2(SO_4)_4$.

| colspan="6" | $K_3Cu_3(Fe_{0.82}Al_{0.18})O_2(SO_4)_4$[15] (Data for ICSD - 67698) |
|---|---|

| colspan="6" | Space group $I2$ (N5): $a$ = 18.667 Å, $b$ = 4.94 Å, $c$ = 18.405 Å, α = 90°, β = 101.5°, γ = 90°, Z=4 |

| colspan="6" | Method[a] – XDS; R-value[b] = 0.12 |

| d($M_i$-$M_j$) (Å) | $Jn$[c] (Å$^{-1}$) | $j^{max\ (d)}$ (Å$^{-1}$) ($\Delta h(X)$[e] Å, $l_n'/l_n$[f], $M_i$-X-$M_j$[g]) | d($M_i$-$M_j$) (Å) | $Jn$ (Å$^{-1}$) | $j^{max\ (d)}$ (Å$^{-1}$) ($\Delta h(X)$[e] Å, $l_n'/l_n$[f], $M_i$-X-$M_j$[g]) |
|---|---|---|---|---|---|
| colspan="6" | *tetrahedron I: O18Cu1Cu2Cu3Fe*[h] |
| d(Cu1-Cu3) 3.220 | J1 -0.0986 | j(O18): -0.0986 (-0.509, 1.8, 122.04°) | d(Fe-Cu3) 2.918 | J4 -0.0761 | j(O17): -0.0400 (-0.169, 1.12, 99.6°) j(O18): -0.0361 (-0.149, 1.29, 98.3°) |
| d(Cu1-Cu2) 2.899 | J2 -0.0756 | j(O17): -0.0331 (-0.139, 1.01, 97.96°) j(O18): -0.0425 (-0.179, 1.00, 99.77°) | d(Fe-Cu1) 3.400 | J5 -0.0671 | j(O18): -0.0671 (-0.385, 1.13, 118.2°) |
| d(Cu2-Cu3) 2.870 | J3 -0.0604 (0.0028) | j(O18): -0.0604 (-0.248, 1.10, 102.38°) j(O12): 0.0632 (0.216, 1.88, 80.89°) | d(Fe-Cu2) 3.364 | J6 -0.0515 | j(O18): -0.0629 (-0.353, 1.13, 116.1°) |
| colspan="6" | *tetrahedron II: O17Cu1Cu2Cu3Fe*[h] |
| d(Cu1-Cu3) 3.331 | J7 -0.0719 | j(O17): -0.0719 (-0.399, 1.04, 118.0°) | d(Fe-Cu2) 3.268 | J9 -0.0854 | j(O17): -0.0854 (-0.455, 1.06, 119.88°) |
| d(Cu2-Cu3) 3.115 | J8 -0.0463 | j(O17): -0.0463 (-0.224, 1.04, 105.9°) | d(Fe-Cu1) 3.193 | J10 -0.0801 | j(O17): -0.0801 (-0.408, 1.05, 116.24°) |
| colspan="6" | *intrachain couplings*[h] |
| d(Cu1-Cu1) 4.940 | $J_b^{Cu1}$ 0.0227 | j(O2): -0.0193 (-0.189, 1.99, 123.5°) j(O4): 0.0395 (0.480, 1.09, 105.4°) | d(Cu1-Cu2) 5.664 | J11 -0.0208 | j(O17): -0.0105 (-0.730, 2.16, 153.1°) j(O18): -0.0103 (-0.725, 2.19, 149.8°) |
| d(Cu2-Cu2) 4.940 | $J_b^{Cu2}$ 0.0276 | j(O5): 0.0240 (0.292, 1.03, 111.2°) | d(Cu1-Cu2) 5.791 | J12 -0.0213 | j(O17): -0.0107 (152.31°) j(O18): -0.0114 (152.76°) |
| d(Cu3-Cu3) 4.940 | $J_b^{Cu3}$ -0.0042 | j(O12): -0.0047 (-0.258, 2.26, 124.6°) | d(Fe-Cu3) 5.552 | J13 -0.0221 | j(O17): -0.0107 (-0.732, 2.23, 148.9°) j(O18): -0.0114 (-0.810, 2.29, 152.0°) |
| d(Fe-Fe) 4.940 | $J_b^{Fe}$ -0.0025 | j(O16): -0.0095 (-0.510, 2.20, 135.4°) | d(Fe-Cu3) 5.918 | J14 -0.0200 | j(O17): -0.0104 (-0.782, 2.16, 153.1°) j(O18): -0.0096 (-0.691, 2.05, 149.8°) |
| colspan="6" | *interchain couplings*[h] |
| d(Cu2-Cu2) 6.333 | J15 -0.0073 | j(O8): -0.0063 (-0.795, 2.98, 174.3°) | d(Cu3-Cu3) 7.722 | J18 -0.0104 | j(O8): -0.0109 (-0.265, 1.95, 144.04°) |
| d(Cu2-Cu3) 6.399 | J16 -0.0018 | j(O8): -0.0018 (-0.293, 3.95, 127.2°) | d(Fe-Cu1) 7.947 | J19 -0.0034 | j(O4): -0.0027 (-0.601, 3.49, 148.3°) |
| d(Cu1-Cu1) 6.805 | J17 -0.0006 | 2x j(O11): -0.0017 (-0.159, 2.07, 135.6°) | d(Cu1-Cu2) 8.567 | J20 -0.0300 | j(O11): -0.0332 (-1.188, 1.25, 174.27°) |

[a] XDS - X-ray diffraction from single crystal.

[b] The refinement converged to the residual factor (*R*) values.

[c] *J*n<0 – AFM, *J*n>0 – FM

[d] *j* - maximal contributions of the intermediate X ion into the AFM component of the *J*n coupling

[e] $\Delta h(X)$ – the degree of overlapping of the local space between magnetic ions by the intermediate ion X.

[f] $l_n'/l_n$ - asymmetry of position of the intermediate X ion relatively to the middle of the $M_i$-$M_j$ bond line.

[g] $M_i$-X-$M_j$ bonding angle

[h] As accepted for *J*n calculations, the Fe position is fully occupied

cations and form two types of oxo-centered tetrahedra: [$O17Cu1Cu2Cu3Fe$] (O17–Cu = 1.91-1.97 Å; O17–Fe = 1.85 Å), and [$O18Cu1Cu2Cu3Fe$] (O18–Cu = 1.79-1.90 Å; O18–Fe = 2.07 Å) with comparatively high strengths of chemical bonds. Let us denote tetrahedra centered with the O17 ion as the type I and O18-centered ones as type II. These oxo-centered [$OCu_3Fe$] tetrahedra are linked through edges (Cu1-Cu2 = 2.899 Å and Cu3-Fe = 2.918 Å) into chains stretched along the *b* axis. The chains are linked to each other through SO$_4$ tetrahedra and K ions. The chains of edge-sharing *Cu1Cu2Cu3Fe* tetrahedra serve as a base of the sublattice of magnetic ions (Fig. 2b, d) in klyuchevskite. Tetrahedra of I and II types alternate in the chain.

Our calculations (Table 2, Fig. 2b, d) demonstrate that strong antiferromagnetic (AFM) couplings exist along all Cu-Cu the tetrahedra edges, except one (Cu2-Cu3 in the tetrahedron of the type I). The main contribution to formation of the AFM character of these couplings is provided by O18 and O17 oxygen ions centering the above tetrahedra. The AFM *J*1 (*J*1 = -0.0986 Å$^{-1}$, d(Cu1-Cu3) = 3.220 Å) coupling is the strongest among them. The antiferromagnetic *J*1, *J*5 (d(Fe-Cu1) = 3.400 Å) and *J*6 (d(Fe-Cu2) = 3.364 Å) couplings in the tetrahedron of the type I and antiferromagnetic *J*7 (d(Cu1-Cu3) = 3.331 Å), *J*8 (d(Cu2-Cu3) = 3.115 Å), *J*9 (d(Fe-Cu2) = 3.268 Å) and *J*10 (d(Fe-Cu1) = 3.193 Å) couplings in the tetrahedron of the type II emerge under effect of O18 and O17 ions, respectively. The contributions to AFM components of interactions *J*2 (d(Cu1-Cu2) = 2.899 Å) and *J*4 (d(Fe-Cu3) = 2.918 Å) along common tetrahedra edges in the chain are provided by both O17 and O18 ions.

The *J*3 coupling (d(Cu2-Cu3) = 2.870 Å) between magnetic Cu2 and Cu3 ions in the tetrahedron of the type I is formed under effect of two oxygen ions (O18 and O12) entering its local space. Although the oxygen O18 ion makes a substantial AFM contribution (*j*(O18) = -0.0604 Å$^{-1}$) to the emergence of the *J*3 coupling, the FM contribution of the O12 ion (*j*(O12) = 0.0632 Å$^{-1}$) exceeds it insignificantly (by 0.0028 Å$^{-1}$) and, thus, makes this coupling weak ferromagnetic. Then, along the zigzag-like chain formed by tetrahedra Cu2-Cu3 edges, the orientation of magnetic moments will be as follows: ↑↑↓↓. Nevertheless, the *J*3 coupling can be hardly considered as a stable one. The point is, the O12 ion is located in the critical position, and a shift from it could result in reorientation of magnetic moments (FM–AFM transition). For example, the decrease of the Cu2-O12 distance from 2.47 Å to 2.34 Å (at the increase of the values of *x* and *z* coordinates of the O12 ion by just 0.01) will yield 5-fold decrease of the FM contribution and induce the transition of the *J*3 coupling into the AFM state. Low accuracy of determination of the crystal structure of klyuchevskite in Ref. 15] by means of X-ray single-crystal diffraction (the refinement converged to the residual factor (*R*) values *R* = 0.12) enables one to assume that the O12 ion has an insignificant role in formation of the *J*3 coupling, so that it is strong and antiferromagnetic (*J*3/*J*1 = 0.61), just like other couplings in tetrahedra.

To sum up, the crystal structure of klyuchevskite causes the emergence of strongly frustrated AFM chains of edge-sharing tetrahedra stretched along the *b* axis, with competing strong nearest neighbor AFM couplings along the tetrahedra edges. In case of reorientation of magnetic moments from antiferromagnetic to ferromagnetic (AFM *J*3 → FM *J*3) along just one of six tetrahedron edges, all the magnetic couplings in the tetrahedron will remain frustrated. In case of complete substitution of Fe$^{3+}$ ions by nonmagnetic Al$^{3+}$ ions at preservation of the crystal structure, the frustrated AFM chain from tetrahedra will transform into a corrugated AFM chain from copper triangles alternately coupled through a side and a vertex (Fig. 2c). This chain will be also frustrated.

Extra intrachain couplings (Table 2, Fig. 2d) at long disgtances *J*11 (*J*11/*J*1 = 0.21, d(Cu1-Cu2) = 5.664 Å) and *J*12 (*J*12/*J*1 = 0.22, d(Cu1-Cu2) = 5.791 Å) are also of the FM type, but they are almost 5-fold weaker than the main ones along tetrahedra edges. The FM couplings $J_b^{Cu1}$ ($J_b^{Cu1}$/*J*1 = -0.23) and $J_b^{Cu2}$ ($J_b^{Cu2}$/*J*1 = -0.28) between similar ions located at a distance of the *b* parameter of the unit cell are slightly stronger. Antiferromagnetic, but very weak couplings $J_b^{Cu3}$ ($J_b^{Cu3}$/*J*1 = 0.04, d(Cu3-Cu3) = 4.94 Å) and $J_b^{Fe}$ (d(Fe-Fe) = 4.94 Å) exist between similar Cu3 and Fe atoms located through the *b* parameter. The AFM character of these couplings indicates to at least twofold increase

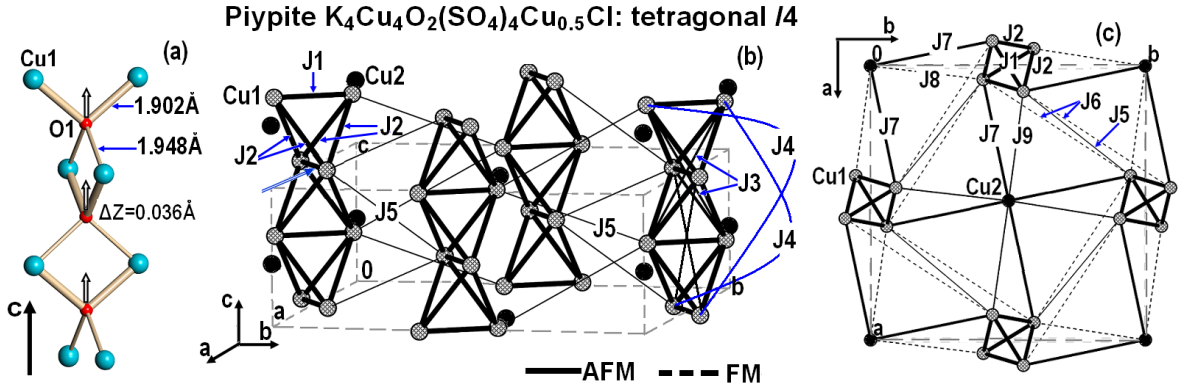

**Fig. 3** Polarization along the *c* axis of the chain of edge-sharing $OCu_4$ tetrahedra in piipite $K_4Cu_4O_2(SO_4)_4MeCl$ (a). The sublattice of magnetic ions $Cu2^{2+}$ formed from chains of edge-sharing $(Cu1_4)$ tetrahedra and the coupling $Jn$ in piipite (b). The interchain $Jn$ couplings in case of occupancy of *Me* positions by magnetic $Cu2^{2+}$ ions (projection of the structure parallel to [001]) (c).

of the *b* parameters of the magnetic unit cell, as compared to that in the crystal unit cell.

The weak AFM couplings $J15$ ($J15/J1 = 0.07$, d(Cu2-Cu2) = 6.333 Å), $J16$ ($J16/J1 = 0.02$, d(Cu2-Cu3) = 6.399 Å), $J17$ ($J17/J1 = 0.006$, d(Cu1-Cu1) = 6.805 Å), $J18$ ($J18/J1 = 0.11$, d(Cu3-Cu3) = 7.722 Å), and $J19$ (d(Fe-Cu1) = 7.947 Å) (Table 2, Fig. 2d) exist between tetrahedral chains. According to the calculations, just one AFM interchain coupling $J20$ ($J20/J1 = 0.34$, d(Cu1-Cu2) = 8.567 Å) is relatively strong. However, the strength of this coupling can be overestimated, since the decrease of coupling strengths accelerates along with the distance increase.

To sum up, our calculations yielded a quasi-one-dimensional spin-tetrahedra system $K_3Cu_3(Fe_{0.82}Al_{0.18})O_2(SO_4)_4$, in which frustrated antiferromagnetic tetrahedra are linked through common edges into chains stretched along the *b* axis. All the couplings between nearest neighbors of copper and iron atoms inside the chain are themselves frustrated due to their location in tetrahedra vertices. Besides, extra competition of these AFM couplings with those between next-to-nearest neighbors Cu and Fe atoms in $J11$-$J10$-$J6$, $J12$-$J8$-$J1$, $J12$-$J9$-$J5$, $J13$-$J10$-$J1$, $J14$-$J7$-$J5$, and $J14$-$J8$-$J6$ triangles exists inside the chain. Antiferromagnetic couplings between chains are very weak.

### 3.2 Piipite (caratiite), $K_4Cu_4O_2(SO_4)_4MeCl$

Piipite $K_4Cu_4O_2(SO_4)_4MeCl$[16] crystallizes in the noncentrosymmetric tetragonal *I*4 system. Just like in klyuchevskite, chains of edge-sharing oxo-centered $OCu_4$ tetrahedra serve as a base of the crystal structure of piipite (Fig. 3a). These chains are stretched in parallel to [001] and linked by $SO_4^{2-}$ and $K^-$ ions into a framework, whose channels contain *Me* and $Cl^-$ ions. However, the nature of the *Me* atoms is not yet established completely. As was assumed in Ref. 16, most probably, the position is half occupied by $Cu^{2+}$, fully occupied by $Na^+$, or occupied by a mixture of the two. Unlike klychevskite, in piipite the chains of edge-sharing $OCu_4$ tetrahedra are polarized. The O1 ions centering tetrahedra are shifted by 0.036 Å (z(O1) = 0.7572) from the center of the $Cu_4$ tetrahedron along the 001 direction (Fig. 3a). As a result, two Cu-O1 bonds are shortened down to 1.902 Å, whereas two others are elongated up to 1.948 Å. If the O1 atom was not shifted (z(O1) = 0.7500), the Cu-O1 distances for all 4 atoms in the tetrahedron would be equal to 1.925 Å.

According to our calculations, the AFM nearest-neighbor $J1$ ($J1 = -0.0719$ Å$^{-1}$, d(Cu1-Cu1) = 2.936 Å) and $J2$ ($J2/J1 = 0.96$, d(Cu1-Cu1) = 3.242 Å) couplings of $Cu_4$ tetrahedra linked through common edges into chains parallel to the *c* axis are the dominating ones (Table 3, Fig. 3b.). The contribution to AFM components of these couplings emerges under effect of oxygen ions (O1) centering the above tetrahedra. Just like $J1$ couplings, the next-to-nearest neighbor $J3$ and $J4$ intrachain couplings are also antiferromagnetic, but 3- and 6-fold weaker, respectively. All these couplings compete to each other, while tetrahedral chains in piipite are frustrated.

**Table 3.** Crystallographic characteristics and parameters of magnetic couplings (*J*n) calculated on the basis of structural data and respective distances between magnetic ions in piypite $K_4Cu_4O_2(SO_4)_4Cu_{0.5}Cl$.

| $K_4Cu_4O_2(SO_4)_4Cu_{0.5}Cl$ [16] (Data for ICSD - 64684) | | | | | |
|---|---|---|---|---|---|
| Space group *I*4 (N79): $a$ = 13.60 Å, $b$ = 13.60.94 Å, $c$ = 4.98 Å, α =90°, β = 90°, γ = 90°, Z=2 | | | | | |
| Method[a] – XDS; R-value[b] = 0.0350 | | | | | |
| d(Cu$_i$-Cu$_j$) (Å) | $Jn$[c](Å$^{-1}$) | $j^{max}$ [d](Å$^{-1}$) ($\Delta h(X)$[e] Å, $l_n$'/$l_n$[f], Cu$_i$-X-Cu$_j$[g]) | d(Cu$_i$-Cu$_j$) (Å) | $Jn$ (Å$^{-1}$) | $j^{max}$ [d](Å$^{-1}$) ($\Delta h(X)$[e] Å, $l_n$'/$l_n$[f], Cu$_i$-X-Cu$_j$[g]) |
| *tetrahedron I:* Cu(1)$_4$ | | | | | |
| d(Cu1-Cu1) 2.936 | $J1$ -0.0719 | $j$(O1): -0.0443 (-0.191, 1.0, 101.04°) $j$(O1): -0.0276 (-0.119, 1.0, 97.79°) | d(Cu1-Cu1) 3.242 | $J2$ -0.0689 | $j$(O1): -0.0689 (-0.362, 1.03, 114.71°) |
| *intrachain couplings in tetrahedral chains* | | | | | |
| d(Cu1-Cu1) 5.781 | $J3$ -0.0211 | $j$(O1): -0.0100 (-0.750, 2.23, 150.75°) $j$(O1): -0.0111 (-0.0111, 2.13, 152.75°) | d(Cu1-Cu1) 4.980 | $J_c^{Cu1}$ 0.0040 | $j$(O2): 0.0028 (0.232, 2.21, 101.97°) |
| d(Cu1-Cu1) 7.753 | $J4$ -0.0117 | $j$(O1): -0.0120 (h=-0.362, 1.02, 150.02°) | d(Cu1-Cu1) 9.960 | $J_{2c}^{Cu1}$ -0.0051 | $j$(Cu1): -0.0131 (-0.650, 1.0, 180°) |
| *interchain couplings* | | | | | |
| d(Cu1-Cu1) 7.240 | $J5$ -0.0011 | $j$(O3): -0.0023 (-0.363, 3.04, 139.2) $j$(O4): 0.0020 (0.462, 4.42, 108.2) | d(Cu1-Cu1) 8.716 | $J6$ 0.0038 | $j$(O3): -0.0126 (-0.422, 1.68, 153.12°) $j$(O5): 0.0131 (0.494, 1.13, 132.90°) |
| *linear chains of Cu2 ions*[h] | | | | | |
| d(Cu2-Cu2) 4.980 | $J_c^{Cu2}$ -0.1460 | $j$(Cl): -0.1460 (-1.810, 1.0, 180°) | d(Cu2-Cu2) 9.960 | $J_{2c}^{Cu2}$ -0.0253 | $j$(Cu2): -0.0131 (-0.650, 1.0, 180°) 2 x $j$(Cl): -0.0061 (-1.810, 3.01, 180°) |
| *couplings between tetrahedral and linear chains*[h] | | | | | |
| d(Cu1-Cu2) 6.169 | $J7$ -0.0619 | $j$(O4): -0.0640 (-1.057, 1.72, 166.36°) | d(Cu1-Cu2) 6.238 | $J9$ -0.0108 | $j$(O2): -0.0089 (-0.795, 2.29, 154.40°) |
| d(Cu1-Cu2) 6.028 | $J8$ 0.0074 | $j$(O5): -0.0128 (-0.0200, 1.75, 133.91°) $j$(O4): 0.0178 (0.306, 1.42, 119.89.14°) | | | |

[a] XDS - X-ray diffraction from single crystal.

[b] The refinement converged to the residual factor (*R*) values.

[c] $Jn<0$ – AFM, $Jn>0$ – FM

[d] *j* - maximal contributions of the intermediate X ion into the AFM component of the *J*n coupling

[e] $\Delta h(X)$ – the degree of overlapping of the local space between magnetic ions by the intermediate ion X.

[f] $l_n$'/$l_n$ - asymmetry of position of the intermediate X ion relatively to the middle of the Cu$_i$-Cu$_j$ bond line.

[g] Cu$_i$-X-Cu$_j$ bonding angle

[h] As accepted for *J*n calculations, the Cu2 position is fully occupied.

Just one coupling $J_c$ ($J_c/J1$ = -0.06, d(Cu1-Cu1) = 4.980 Å) between copper ions in the chain located across the *c* parameter is very weak and ferromagnetic, but also competes with the weak AFM $J_{2c}$ ($J_c/J1$ = 0.07) coupling that is the next along the *c* parameter. Chains are located at large distances from each other. The nearest between chains $J5$ ($J5/J1$ = -0.015, d(Cu1-Cu1) = 7.240 Å) coupling is very weak antiferromagnetic. The subsequent couplings at distances within 9 Å are weak ferromagnetic.

If one assumes that the *Me* position (2a: x = 0, y = 0, z = 0.448) is at least half-occupied by the magnetic Cu(2)$^{2+}$ ion, then strong AFM $J7$ ($J7/J1$ = 0.86, d(Cu1-Cu1) = 6.169 Å) and weaker AFM $J9$ ($J9/J1$ = 0.15, d(Cu1-Cu1) = 6.238 Å) and FM $J8$ ($J8/J1$ = -0.10, d(Cu1-Cu1) = 6.028 Å) couplings will emerge between frustrated tetrahedral chains and Cu2 ions.

If the *Me* position was fully occupied with magnetic $Cu2^{2+}$ ions, then the piipite structure would consist of two types of chains stretched along the *c* axis: linear chains -Cu2-Cu2- of strong AFM $J_c^{Cu2}$ couplings formed due to intermediate $Cl^-$ ions and frustrated chains of AFM tetrahedra $Cu1_4$ (Table 3, Fig. 3c). These chains are linked to each other through strong AFM *J*7 couplings.

However, coordination in the form of a virtually regular octahedron is not characteristic of the $Cu^{2+}$ ion. It is more similar to the coordination of the nonmagnetic $Cu^{1+}$ ion, if one considers 2 collinear axial Cu-Cl bonds (2.49 Å) as shortened ones for large chlorine ions and, in addition, 4 elongated Cu-O5 bonds (2.50 Å) with small oxygen ions in the octahedron equatorial plane.

To sum up, the results of calculations of characteristics of magnetic couplings by the crystal chemistry method and the analysis of their competition in the structures of noncentrosymmetric minerals klyuchevskite ($K_3Cu_3(Fe_{0.82}Al_{0.18})O_2(SO_4)_4$) and piipite ($K_4Cu_4O_2(SO_4)_4Cu_{0.5}Cl$) assume that their magnetic structures comprise quasi-one-dimensional systems. The magnetic structures of these minerals are formed by dominating in strength antiferromagnetic chains of edge-sharing tetrahedra. All the couplings in chains are frustrated. Antiferromagnetic and ferromagnetic couplings between chains are very weak.

Besides, in piipite the strong frustration of magnetic interactions (absence of magnetic ordering) is combined with the presence of electric polarization in tetrahedra chains. The O1 ions centering $Cu_4$ tetrahedra are shifted from the tetrahedron center along the 001 direction (Fig. 3a). This polarization can be also considered as the shift along the 00-1 direction of the chain of tetrahedra of $Cu^{2+}$ cations relatively to $O^{2-}$ anions centering these tetrahedra. Our calculations demonstrate that the shift of the O1 ion to the tetrahedron center (at preservation of acceptable Cu-O bond lengths) will yield substantial changes in neither strength nor character of magnetic couplings, since the position occupied by the O1 ion is not a critical one.

### 3.3. Uniqueness of tetrahedral spin chains in klyuchevskite and piipite

Magnetism of tetrahedral spin chains is of great interest for theoretical and experimental studies. In the literature, zigzag-like chains of corner-sharing tetrahedral [2-6] (Fig. 4a), the tetrahedral-cluster spin chain [48-50] (Fig. 4b), and the chain of edge-sharing tetrahedral [51, 52] (Fig. 4c) are considered as quasi-one-dimensional frustrated spin-tetrahedral systems.

Among these three tetrahedral quasi-one dimensional spin systems, the only widely spread one contains corner-sharing tetrahedra chains, for instance, in $Cu_3Mo_2O_9$ [53, 54] and in pyrochlore lattices [30]. We did not manage to find in the literature any data on experimental studies of magnetic compounds, in which spin tetrahedra would be linked into chains via edges.

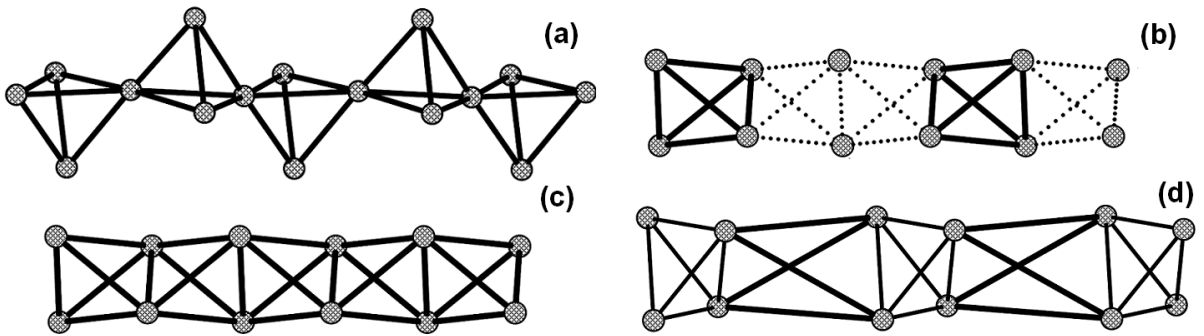

**Fig. 4** Schematic representation of quasi-one dimensional frustrated spin-tetrahedral systems: (a) zigzag-like chains of corner-sharing tetrahedra, (b) tetrahedral-cluster spin chain, (c) chain of edge-sharing tetrahedra, and (d) chains in $Cu_2Te_2O_5X_2$.

Instead of the tetrahedral-cluster spin chain and the spin chain of edge-sharing tetrahedra, the system $Cu_2Te_2O_5X_2$ with $X$ - Cl, Br has been examined in the literature [36, 37, 56, 57]. It contains tetrahedral clusters of copper ions ($Cu_4$), which are not edge-sharing, but align to tubes or chains. Strong inter-tetrahedral couplings are present in such chains. However, in spite of numerous theoretical and experimental studies, the accurate dimensionality of the $Cu_2Te_2O_5X_2$ system is not clear. There exists the opinion that this system is three-dimensional rather than quasi-one dimensional, since interchain interactions in it are rather strong1 [7, 37, 56, 57]. Different models of the magnetic state of this system were examined in theoretical terms. The authors of Refs. 49-51 investigated frustration in the tetrahedral-cluster spin chain and the chain of edge-sharing tetrahedra within the frames of the model of two-leg spin ladders. Kotov et al. [58, 59] demonstrated the role of antisymmetric Dzyaloshinsky-Moriya (DM) [60, 61] spin-spin interactions in inducing weak antiferromagnetism in $Cu_2Te_2O_5Br_2$.

We found frustrated AFM spin chains of edge-sharing tetrahedra (Fig. 4b) in two minerals: klyuchevskite and piypite. The uniqueness of these minerals consist not only in the very presence of such spin tetrahedra chains in them, but also in the fact that the AFM character and frustration of exchange interactions in these chains are caused by oxygen ions centering copper tetrahedra ($OCu_4$). Since the shift of these oxygen ions is limited by small sizes of $Cu_4$ tetrahedra, reorientation of magnetic moments (AFM → FM) along the tetrahedra edges and, therefore, the frustration suppression due to changes in the character of exchange interactions in the chain will be impossible.

## 4 Conclusions

As was demonstrated by the calculations of the sign and strength of magnetic interactions based on the structural data, noncentrosymmetric minerals klyuchevskite ($K_3Cu_3(Fe_{0.82}Al_{0.18})O_2(SO_4)_4$) and piipite ($K_4Cu_4O_2(SO_4)_4MeCl$) were frustrated quasi-one-dimensional spin-1/2 tetrahedral systems containing AFM spin chains of edge-sharing $Cu_4$ tetrahedra. The magnetic system of such compounds is disordered because of frustration of strong exchange interactions in tetrahedra, in which antiparallel orientation of all the nearest neighbors is impossible due to geometric reasons. At the same time, in the crystal structure of piipite, one detects the existence of electric ordering (polarization) of chains of $OCu_4$ tetrahedra, since the oxygen O1 ions that center them are shifted from the tetrahedra centers along the 001 direction.

In conclusion, one should mention some problems inherent to determination of the crystal structure of these minerals, but cannot be solved by crystal chemistry methods. Although the existence of frustration of magnetic interactions in these minerals is caused by their crystal structures and raises no doubts, it is still unclear whether their magnetic systems will remain completely disordered until the temperature absolute zero (will become the quantum spin liquid). In view of this, it appears of interest to reveal the possibility of having an essential role in magnetic ordering for very weak long-range AFM interactions linking base elements of the magnetic structure (frustrated tetrahedra chains with strong AFM interactions). Another problem consists in the possibility of the emergence of the spatially modulated spin structure in these minerals as a result of activation of forces of the relativistic nature (of Dzyaloshinskii–Moriya) [60, 61]. The crystal structures of klyuchevskite and piipite are favorable for this, since their space groups ($I2$ and $I4$, respectively) do not contain the inversion center, symmetry planes, and rotoinversion axes.

**Acknowledgments** The work was partially supported by the Program of Basic Research "Far East" (Far-Eastern Branch, Russian Academy of Sciences), project no. 15-I-3-026.